\documentclass[12pt]{article}
\usepackage{graphicx}

\newcommand{\ug}{\; = \;}

\newcommand{\ubf}{\mbox{\boldmath $u$}}
\newcommand{\Ubf}{\mbox{\boldmath $U$}}

\newcommand{\Ebf}{\mbox{\boldmath $E$}}

\newcommand{\pbold}{\mbox{\boldmath $p$}}
\newcommand{\Hbf}{\mbox{\boldmath $H$}}

\newcommand{\xbf}{\mbox{\boldmath $x$}}

\newcommand{\text}{\rm}

\newcommand{\Scal}{{\cal S}}

\newcommand{\bb}{\begin{equation}}
\newcommand{\ee}{\end{equation}}
\newcommand{\bega}{\begin{eqarray}}
\newcommand{\ega}{\end{eqnarray}}
\newcommand{\begae}{\begin{eqnarray*}}
\newcommand{\egae}{\end{eqnarray*}}

\newcommand{\1}{1\!\!1}

\newcommand{\h}{\hspace*{4ex}}

\newcommand{\cent}{\centerline}
\newcommand{\vs}{\vspace*}

\newcommand{\Lcu}{\mathcal{L}\stackrel{^\uparrow}{_+}}
\newcommand{\Lcd}{\mathcal{L}\stackrel{^\downarrow}{_+}}

\begin{document}

\baselineskip 0.8cm

\begin{center}

{{\large {\bf On the ``Non-Restricted special Relativity" theory (NRR) and further comments on ``Cherenkov vs X-waves"}}}

\end{center}

\vs{5mm}

\cent{ Michel Zamboni-Rached, }

\vs{2mm}

\centerline{{\em DMO, Facultade de Engenharia El\'etrica, UNICAMP,
Campinas, SP, Brazil.}}

\vs{5mm}

\cent{ Erasmo Recami, }

\vs{2mm}

\cent{{\em INFN---Sezione di Milano, Milan, Italy,} {\rm and}}
\cent{{\em Facolt\`a di Ingegneria, Universit\`a statale di Bergamo,
Bergamo, Italy.}}

\vs{3mm}

\centerline{\rm and}

\vs{3mm}

\cent{ Ioannis M. Besieris }

\vs{2mm}

\cent{{\em The Bradley Deptartment of Electrical and Computer Engineering}}
\cent{{\em Virginia Polytechnic Institute and State University}}
\cent{{\em Blacksburg, VA-24060, USA.}}

\vs{16mm}

{\bf Abstract ---}  Our aim in this paper is to recall some essential points of ``Extended special Relativity", now more
correctly called ``Non-Restricted special Relativity" theory (NRR), and in particular of the Extended Maxwell Equations;
as well as to set forth some further comments on the {\em basic} differences
between Cherenkov Radiation and the so-called X-shaped Waves, met within the more recent realm of the
Non-diffracting Waves (also known as Localized Waves). \ The occasion is furnished by some very recent Seshadri's comments[1]
on a previous article of ours, titled ``Cherenkov radiation versus X-shaped localized waves" (see[2],
and arXiv:0807.4301[physics.optics]), and not less on NRR itself.\\

OCIS codes: 320.5550; 350.7420; 070.7345; 350.5500; 070.0070;
100.7410; 050.050; 000.1600; 000.2690; 000.6800; 250.5530; 260.0260.\hfill\break
PACS nos.: \ 41.60.Bq; 03.50.De; 03.30.+p; 41.20;Jb;
04.30.Db; 42.25.-p;  42.25.Fx; 47.35.Rs.\hfill\break

\

\

\

\

{\Large{ \bf 1. \ General observations}}

\

Our aim in this paper is to recall some essential points of ``Extended special Relativity", now more
correctly called ``Non-Restricted special Relativity" theory (NRR), and in particular of the Extended Maxwell Equations;
as well as to set forth some further comments on the {\em basic} differences
between Cherenkov Radiation and the so-called X-shaped Waves, met within the more recent realm of the
Non-diffracting Waves (also known as Localized Waves). \ The occasion is furnished by some very recent Seshadri's comments[1]
on a previous article of ours, titled ``Cherenkov radiation versus X-shaped localized waves" (see[2],
and arXiv:0807.4301[physics.optics]), and not less on NRR itself.

\h Let us first answer Seshadri's comments[1] to the previous article
of ours, titled ``Cherenkov radiation versus X-shaped localized waves"[2].

\h Actually, as we were saying, Seshadri's comments do not refer too much to our article[2],
whose aim was mainly showing in detail that X-shaped localized waves have
nothing to do with Cherenkov radiation, as maintained by contrast in
Ref.[3]. It should be pointed out that the so-called Localized Waves (LW) are
nondiffracting (``soliton-like") solutions to linear wave equations,
and a host of theoretical, mathematical, numerical-simulation, and
experimental works have demonstrated them to exist with subluminal, luminal or
superluminal peak-velocities $V$ (see, e.g., [4] and references therein). These wave
packets are remarkable and have been thoroughly investigated more for their limited-diffraction
and self-reconstruction properties than for their group-velocities. \ However, merely
for mathematical  and experimental[5,6] reasons, the ones that drew more attention
were the superluminal ``X-shaped" ones, which in fact started to be considered even for
practical applications in 1992. The X-waves have the shape of a double cone. For this
reason, some authors have been tempted  ---let us confine ourselves to electromagnetism--- to look
for links between them and the Cherenkov radiation: a link that we have throroughly shown in [2]
to be untenable.

\h The comments in Ref.[1] are {\em partially} related with our article [2]; but in reality they
appear to be more addressed to a previous paper of ours, with the rather
different title  (and subject) ``Localized X-shaped field generated by a superluminal charge"[7]:
A very different subject, indeed, since even the superluminal LWs are known[4] to consist of
soperpositions of ``lateral" (feeding) waves, which travel at the ordinary speed of light[8].
Actually, in our article [2] we were induced to go briefly back to questions faced by us
in Ref.[7], and to restate one of those questions more rigorously in terms of half-retarded
and half-advanced Green functions,
{\em only} because a part of Ref.[3] itself was inadvertently pointing in that direction.

\h The criticisms in Ref.[1] regard moreover the {\em theory of Extended Special Relativity,
correctly called by Seshadri Non-Restricted Special Relativity\/} (NRR), which, on the basis
of the ordinary postulates of Special Relativity (SR), chosen of course in a
modern way, had shown how the theory of SR in its Einstein-Minkowski formulation could
have predicted already in 1908 the existence of antiparticles, on one hand, as well as
described superluminal motions without severe violations or paradoxes, on the other
hand.  Incidentally,
superluminal motions were investigated in pre-relativitic times by J.J.Thomson (1889),
O.Heaviside (1892), A.Des Coudres (1900), and particularly, in 1904 and 1905, by
Sommerfeld[9,10]. A brief resume of the very stimulating results about a radiating superluminal charge
determined, e.g., by Sommerfeld
can be found in [11]. In post-relativistic times,
the construction of a NRR started with Sudarshan {\em et al.}'s papers[12,13], and was continued
by one of the present authors and coworkers (see, e.g., review [14] and refserences therein).
For future use, let us summarize some characteristics[14] of NRR, an interesting theory that allows
a deeper understanding of physics (cf., e.g., Refs.[15,16]),as we shall see, even if tachyons
would not exist as asymptotically free objects in our cosmos. \  A modern choice of the postulates
of SR is known to be: (i) Principle of relativity;  (ii) space-time homogeneity and space
isotropy. Form these two postulates it has been demonstrated since 1911 that the existence of
one, and only one, {\em invariant} speed follows; and experience tells us it to be the
speed $c$ of light in the vacuum.  As a consequence, following, e.g., Landau, one gets

\bb ds'^2 \ug \pm ds^2 \label{eq1} \ee

$ds^2$ being the square of the spacetime distance element, that is, the four-dimensional
quantity $ds^2 \equiv dx_\mu dx^\mu \equiv dt^2 - d\xbf^2$, where $d\xbf^2 \equiv dx^2 +
dy^2 + dz^2$. By choosing the sign plus (+), one describes the subluminal world, while
the superluminal one is described[14] when choosing the sign minus (-). Let us notice,
therefore, that also in NRR the $ds^2$ is invariant, except for a sign. More generally, all
the quadratic forms are analogously invariant: for instance, in the simplest
cases, $x'_\mu x'^\mu = \pm x_\mu x^\mu \; ;  \ p'_\mu p'^\mu = \pm p_\mu p^\mu \; ; \
x'_\mu p'^\mu = \pm x_\mu p^\mu$; \ the minus sign entering into play when
passing on from a subluminal object (bradyon, B) to a superluminal object (tachyon, T).
It is on the basis of Eqs.(\ref{eq1}), and of some obvious further assumptions,
that the ordinary subluminal Lorentz transformations (LT) are written down in correspondence
with the sign +, and analogously
the superluminal Lorentz ``transformations" (SLT) in correspondence with the sign -. We shall
come back again to this point.

\

\h {\em Let us forget for a moment about tachyons,} and consider only the sign + in Eqs.(\ref{eq1}),
with a subluminal boost along $x$. Since conditions (\ref{eq1}) are quadratic, each Lorentz
transformation $L$ can
be taken with a double sign [this is another double sign yielded by mathematics: which
has nothing to do with the one in Eqs.(\ref{eq1})].  A little more formally, let us recall
that the set of all subluminal LTs consists of four pieces, which form a noncompact, non-connected
group (the Full Lorentz Group). Wishing to confine ourselves to spacetime ``rotations" only, i.e., to
the case det $L=+1$, we are left with two pieces only: \ $\{L\stackrel{^\uparrow}{_+}\}$, \ and \
$\{L\stackrel{^\downarrow}{_+}\}$, \ which give origin to the group of the proper (orthochronous
\emph{and} antichronous) transformations \  $\mathcal{L}_+ \equiv \mathcal{L}\stackrel{^\uparrow}{_+} \cup
\mathcal{L}\stackrel{^\downarrow}{_+} \equiv \{L\stackrel{^\uparrow}{_+}\} \cup \{L\stackrel{^\downarrow}{_+}\}$,
and to the subgroup of the (ordinary) proper orthochronous transformations $\mathcal{L}\stackrel{^\uparrow}{_+}
\equiv \{L\stackrel{^\uparrow}{_+}\}$. In other words, $\mathcal{L}_+$ can be written as

\begin{equation}
\mathcal{L}_+ = \Lcu \otimes Z(2) \quad ; \quad Z(2) \equiv
\{\sqrt{+1}\} \equiv \{+1,-1\} \; . \label{eq2}
\end{equation}

Let us skip in the following, for simplicity's sake, the subscript
$+$. \ Given an orthochronous transformation $\overline{L}^\uparrow \in \Lcu$, another antichronous transformation
$\overline{L}^\downarrow$ always exists such that \ $\overline{L}^\downarrow = (- 1) \overline{L}^\uparrow$, \ for all
$\overline{L}^\downarrow \in \Lcd$, \ and vice-versa. Such a one-to-one correspondence does obviously allow one
to write $\Lcd = -\Lcu$.
Usually, even the latter piece is discarded, since the antichronous LTs change the time direction
sign (something that is not acceptable).  But any LT acts also on the dual four-dimensional space, that is, on the
energy-momentum space, changing the energy sign too (something not acceptable as well). {\em But} it has been shown long
time ago, in a large number of papers (see, e.g., Refs.[12-14,17] and refs. therein), that those
two paradoxical occurrences, when they are ---as they actually are--- simultaneous, lead
necessarily to the conclusion that the antichronous LTs describe the corresponding
antiparticles (travelling of course forward in time, with positive energy), an orthodox
conclusion, that could have been inferred in 1908, and that has been confirmed by
the experimental discovery of antiparticles.  Namely, the theory of SR, once
based on the whole proper Lorentz group and not only on its
orthochronous part, describes a Minkowski spacetime populated
by both matter and antimatter.

\

\h What is more important for us is that, {\em going now back to the superluminal case} [minus sign in
Eqs.(\ref{eq1})], also the SLTs are to be taken with a double sign. The inverse
transformations do once more exist, contrary to a claim in Ref.[1]. The situation in this case is however more complex,
since, e.g., two successive SLTs yield an ordinary LT; actually, the group $\mathcal{G}$ of all
(subluminal or superluminal) transformations can be formally written, with \ $Z(4) \equiv
\{ \sqrt[4]{+1} \}$, \ as

\begin{equation}
\mathcal{G} = \Lcu \otimes Z(4) \; . \label{eq3}
\end{equation}

We shall come back again to this point too. \ In any case, the Stueckelberg-Feynman-Sudarshan
(SFSR) switching principle, invoked before, does work also in the case of SLTs, even if in this
case the distinction particle/antiparticle is no longer Lorentz invariant[12-14]. See the two-sheeted hyperboloids
(case of Bs), and the one sheeted hyperboloid (case of Ts), depicted for instance in figure 5 of Ref.[14],
and here reproduced as our Fig.1.  In this figure, it is the symmetry with respect to the plane $E=0$
which leads from particles to antiparticles (namely, from Bs to anti-Bs in the two-sheeted case, and
from Ts to anti-Ts in the one-sheeted case). Let us add that the SFSR reinterpretation procedure not only
can, but {\em must} be applied[12-14,17].

\begin{figure}[!h]
\begin{center}
 \scalebox{2.0}{\includegraphics{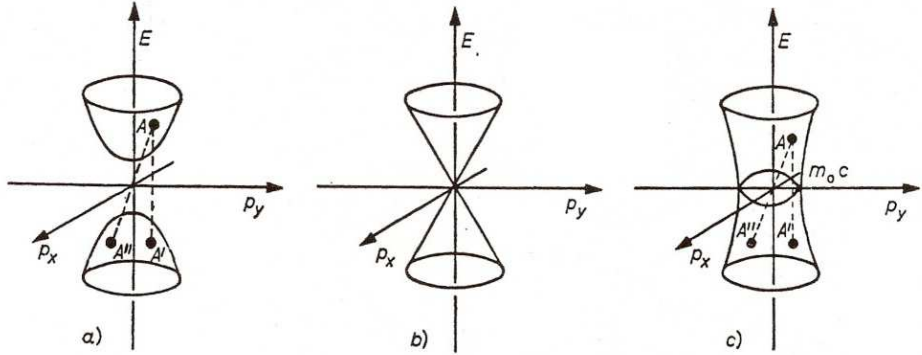}}
\end{center}
\caption{Pictures (in 3D only) of the surfaces $p^2 \equiv E^2 - \pbold^2 = \pm m_0^2$: In figure a) for bradyons,
when $p^2 > 0$; \ in b) for luxons, when $p^2=0$; \ and in c) for tachyons, when $p^2 < 0$. \ The symmetry matter/antimatter[14]
is the symmetry w.r.t. the plane $E=0$: See the text. }
\label{fig1}
\end{figure}

\h Below, we shall need another result of NRR. By implementing the aforementioned rule that SLTs change the
sign of all the quadratic forms, in Refs.[18,14] it was demonstrated that, if an object is a sphere
(or just a point) when at rest, it will appear, when traveling with superluminal speed $V$, as the region
contained between an indefinite double-cone (with semi-angle $\alpha$ given by the simple relation $\tan \alpha = \sqrt{V^2-1})$
and an internal two-sheeted hyperboloid (or just as a double cone): See Fig.2, reproduced from Refs.[18,14].
Notice that this holds, even if the speed $c$ of light is a limiting speed, which in both SR and NRR cannot be crossed,
neither from the left nor from the right[14].

\begin{figure}[!h]
\begin{center}
 \scalebox{1.8}{\includegraphics{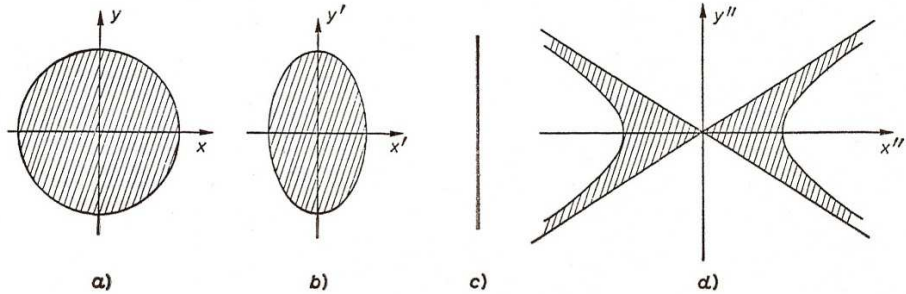}}
\end{center}
\caption{An intrinsically spherical (or pointlike, at the limit)
object appears in the vacuum as an ellipsoid contracted along the
motion direction when endowed with a speed $v<c$. \ By contrast,
if endowed with a speed $V>c$ (even if the $c$-speed barrier
cannot be crossed, neither from the left nor from the right), it
would appear[14,18] no longer as a particle, but as
occupying the region delimited by a double cone and a two-sheeted
hyperboloid ---or as a double cone, at the limit--, and moving
with
Superluminal speed $V$ [the cotangent square of the cone
semi-angle, with $c=1$, being $V^2-1$. For simplicity, a space axis is
skipped. \ This figure is taken from our Refs.[14,18]. }
\label{fig2}
\end{figure}

\

\

\

\

{\Large{ \bf 2. \ Specific replies}}

\

\

{\bf 2.1 \ Reply to Seshadri's Section 2}

\

The Wheeler-Feynman approach --just optional in the case of photons--
becomes necessary in the case of tachyons, so that one has to consider
half advanced and half retarded potentials, as we did, e.g., in Subsection 2D
of Ref.[2] (and in [7]). Then, one gets that a superluminal charge does not emit
Cherenkov radiation in the vacuum (as it must obviously be the case in classical
physics, since that radiation is induced by the charge from a medium).

\h In performing the integration in the complex $\omega$ plane, with $\zeta \equiv z-vt$, Seshadri[1]
goes on to $\omega_r + i \; \omega_i$, so that his integration is essentially along the real
axis, but indented from above at the two existing poles whenever the imaginary part $\omega_i$ is assumed
to be positive;
afterwards, by closing the integration contour in the upper plane, he gets of course that for $\zeta > 0$
the integral yields zero. {\em However,} if one chooses a negative $\omega_i$, the integration in
the upper part of the complex plane ($\zeta > 0$) does not yield zero! \  The choice depends on the imposed
physical conditions, ours being that the superluminal charge should not radiate when
traveling with constant speed in the vacuum; for physical reasons already mentioned in 1973 in old papers
like [19] [but indeed discussed among the ``superluminal" scholars (like E.C.G. Sudarshan, and Recami)
since 1967]. Cf. figure 2 of Ref.[2], here reproduced as Fig.3, depicting how the total energy flux crossing a
large spatial surface containing the charge itself is just zero.  Our aforementioned choice has been implemented[2,7] by adding half
retarded and half advanced Green function: which implies the existence of both the rear and the front cones.
[Incidentally, Seshadri's choice (unacceptable for physical reasons) refers in a sense to a limiting case
of the initial value problem discussed in Ref.[20].]

\begin{figure}[!h]
\begin{center}
 \scalebox{1.9}{\includegraphics{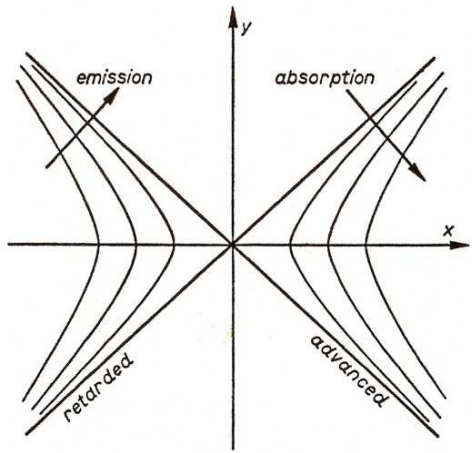}}
\end{center}
\caption{The spherical equipotential surfaces of the electrostatic field
created by a charge at rest get transformed into two-sheeted
rotation-hyperboloids, contained inside an unlimited double-cone, when the
charge travels at superluminal speed[14]. This figures shows,
among the others, that a Superluminal charge traveling at constant speed,
in a homogeneous medium like the vacuum, does {\em not} lose energy. See the text.}
\label{fig3}
\end{figure}

\h Let us mention, incidentally, that interesting results have been found even
in the case of a supersonic point source traveling in a fluid at rest, with respect to
a static observer[21,22].

\h Let us add that one must have recourse to the Wheeler-Feynman approach, since,
when in the presence of a tachyon, {\em laws} such as Retarded Causality are required
to be covariant by the Principle of Relativity, but {\em description details}
as the emitter/absorber labels do not have to be (and are not) invariant[14,17].
This is a known consequence of the very SFSR reinterpretation procedure, which
enters the play when solving[14,17,23] the so-called causal paradoxes (for a thorough
solution of the causal paradoxes associated with tachyonic motions, see Ref.[17]).

\h Before going on, let us stress that the Non-restricted Theory of Special Relativity
allows describing also antimatter and superluminal motions: But this, as already
mentioned, does not mean {\em a priori} that tachyons must exist as asymptotically free
objects in our cosmos; for reviews of the actual experimental possibilities, one
could look, e.g., at Refs.[24,25].

\

{\bf 2.2 \ Reply to Seshadri's Section 3}

\

This is a correct criticism, in the sense
that, because of a misprint, we wrote down a wrong sign for the result of the contested integral.
But our minus (instead of plus) sign does not influence outgoing or
incoming waves since the physically meaningful quantities are obtained
through the Poynting vector, which does not depend on that sign.

\

\

\

\

{\bf 2.3 \ Reply to Seshadri's Section 4}

\

As mentioned above, the general Lorentz transformations (GLT)
form a group, and the inverse transformations
exist always[14]. The SLTs form a group not by themselves, but
together with the ordinary LTs: Let us repeat that, in fact,
two successive SLTs constitute an ordinary LT,
for physical reasons easily understandable on the
basis of NRR, and in particular of its ``duality
principle" (cf., for example, Sec.5 in Ref.[14], which is downloadable for instance
from the site www.unibg.it/recami).

\h The transformations (6) and (7) of Ref.[1], invented
by its author, are not acceptable as SLTs: actually, it can
be easily verified that they imply \ $ds'^2 = + ds^2$, \ which
refers to the case of the ordinary (subluminal) LTs. \ Indeed, the
fundamental relation (\ref{eq1}), as we said, reads

\begin{displaymath}
ds'^2 \ug \left\{
\begin{array}{ll}
+ds^2 \ \ \ \ \ \ \ \ \ {\rm for} \ \ {v^2 < c^2 \; ;}\\
-ds^2 \ \ \ \ \ \ \ \ \ {\rm for} \ \ {v^2 > c^2 \; .}
\end{array}
\right.
\end{displaymath}

All the following considerations in [1], therefore, are
potentially incorrect. In
particular, as we have already seen, the SLTs do admit an inverse transformation[14].
We are grateful, nevertheless, to Seshadri for the attention he
kindly paid to our SLTs[14], and to our extension of Maxwell
equations.  \ Indeed, the calculations in Sec.4 of Ref.[1], relative to the
Maxwell equations, are in the right direction, and reproduce the generalized Maxwell
equations, written by Mignani and Recami at the beginning of the seventies, and
rewritten, e.g., as Eqs.(205) in the 1986 Ref.[14].

\h Since these are rather interesting results, let us seize the present opportunity
for recalling our point with respect to the Maxwell equations. Let us first of all
observe that the electric charge ought to be actually called ``electromagnetic charge" since it creates also a
magnetic field as soon as it moves. Then, it loses meaning
to look for magnetic charges, that is, for ordinary magnetic
poles, which would be inconsistent with the Universality
of the electromagnetic interactions. The Maxwell equations would
however remain (inelegantly) incomplete and non-symmetric.
The solution in principle is the following: If one insists in
calling electric the subluminal charge ---which, in suitable
units, creates more electric that magnetic field components---, then
you can call magnetic the superluminal charge ---which creates
more magnetic than electric field components (see figure 46 of Ref.[14], here
reproduced as our Fig.4). \ Indeed, the
superluminal charges happen to contribute to Maxwell equations
just in the places where a contribution was expected from magnetic
monopoles.

\begin{figure}[!h]
\begin{center}
 \scalebox{2.2}{\includegraphics{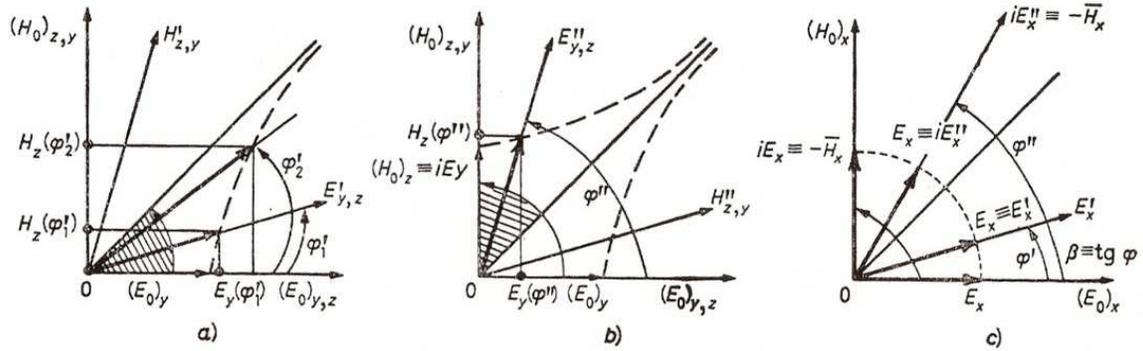}}
\end{center}
\caption{Let us consider in a sistem af rest a purely electric uniform field $\Ebf$ parallel, e.g., to $y$. When moving along $x$
with respect
to that system, one observes also a magnetic field along $z$: Figure a) depicts the ordinary subluminal case with $v\equiv v_x<c$,
in which case
(in Heaviside-Lorentz units) it is $H_z < E_y$.  When moving superluminally with $V\equiv V_x>c$, the magnetic field $H_z$ becomes[14]
(in the same units) larger than the electric field $E_y$: See Figure b); and when $V \rightarrow \infty$ we are left with a
purely magnetic field[14]. \ See the text.}
\label{fig4}
\end{figure}

\h Going back to the form of the SLTs, let us add that the fundamental theorem of
NRR is [with $u^2<1; \ U^2 \equiv 1/u^2 >1$; and $c=1$]:

\begin{equation}
{\rm SLT}(\Ubf) = \pm \Scal \cdot {\rm LT}(\ubf) \; , \label{eq4}
\end{equation}

where the velocities $\Ubf$ and $\ubf$ are parallel, and[14] \
$\Scal \equiv \Scal_4 = i \1$. \ In 2D the theory becomes very
simple, rather interesting and rich of useful consequences[14,26]
(at the extent that somebody believed it to be so elegant, as to be necessarily
true...).  In 4D, however, the
best mathematical expression for the SLTs is an open problem (see subsections
14.1--14.7 of Ref.[14]). In fact, our Fig.2 shows that SLTs should
transform points into double cones, that is, manifolds into
manifolds. Anyway, Barut and Chandola[27], reasoning in terms of
mere analytical continuations, adopted for the SLTs in 4D a (very formal)
expression of the type [$c=1; \ U^2 > 1; \ \gamma \equiv 1 / \sqrt{U^2 - 1}$]:

\begin{equation}
 \pm \xbf' \ug i \xbf + {{\gamma - 1} \over U^2} \Ubf (\Ubf \cdot \xbf) - \gamma \Ubf t \; ; \ \ \ \ \
\pm t' \ \ug  \gamma (t - \Ubf \cdot \xbf ) \; , \label{eq5}
\end{equation}

without meeting any difficulties.  We shall come back to this problem
in our next Subsection, when replying to Seshadri's Sec.5.

\

\

{\bf 2.4 \ Reply to Seshadri's Section 5}

\

Let us immediately say that Eqs.24 and 25 in Ref.[1] are not at all
the  ``original superluminal Lorentz transformations"
characterizing NRR, as claimed by the author, who does not seem to be aware of
Ref.[14]. These equations are patently not Lorentz
covariant, because they do not satisfy Eq.(1). (Only in the 2D case
they result to be not far from the correct ones, which are already known[14,26]). \ Incidentally, such unsatisfactory
relations, supposed to be SLTs, were proposed several times in a number of papers that appeared in the seventies
and eighties, and were regularly confuted.

\h It may be somewhat odd to see naive, wrong equations like (24, 25) of Ref.[1]
associated with NRR, a theory that, even if not yet perfect from the {\em formal} point of view, has been
(thoughtfully)
developed along decades, from 1962 (Ref.[12]) to at least 1986 (Ref.[14]).

\h Let us repeat, first of all, that in NRR each transformation admits its inverse transformation[14],
as already mentioned. \ But we have to discuss a more
important issue about SLTs, by referring ourselves for simplicity to a superluminal
boost along $x$.

\h Fig.2 is sketched in Fig.5. It indicates a characteristic of NRR, namely that, if
the initial subluminal object has sizes $\Delta x, \ \Delta y$ and $\Delta z$ (determined by
its intersections with the space axes), the corresponding superluminal object moving
along $x$ with speed $V$ has
along $x$ the size $\Delta x' = \Delta x \sqrt{V^2 -1}$, regularly given by the generalized Lorentz
contraction formula[14]. But it does not have real intersections with the transverse Cartesian axes,
so that[14] (forgetting here for the double sign) it is endowed with
$\Delta y' = i \; \Delta y$; and $\Delta z' = i \; \Delta z$.  We see how in the transverse
space directions one appears to be formally in need of imaginary units ---as implied by our previous Eqs.(3) and (4)---
to represent the fact that (passing on to the limiting case) a point is transformed into a double cone.
In any case, the SLTs, and the GLTs, provided by
NRR, can be found for instance in pages 123--126 of Ref.[14], while the elegant SLTs in 2D can be
found, e.g., in pages 26--30 of the same review [14] (or in Ref.[26]).

\begin{figure}[!h]
\begin{center}
 \scalebox{1.6}{\includegraphics{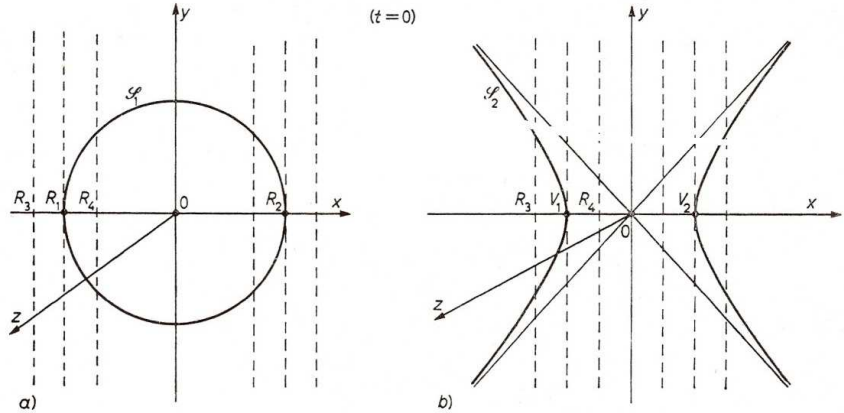}}
\end{center}
\caption{Sketch of Fig.2 showing that, if
the initial subluminal object has sizes $\Delta x, \ \Delta y$ and $\Delta z$ (determined by
its intersections with the space axes), the corresponding superluminal object moving
along $x$ with speed $V$ has
along $x$ the size $\Delta x' = \Delta x \sqrt{V^2 -1}$, regularly given by the generalized Lorentz
contraction formula[14]. {\em But} it does not have real intersections with the transverse Cartesian axes,
so that[14] (forgetting here for the double sign) it is characterized by
$\Delta y' = i \; \Delta y$; and $\Delta z' = i \; \Delta z$.  See the text.}
\label{fig5}
\end{figure}

A further observation is the following: From Fig.5, and Eq.(3), one realizes the remarkable fact that a SLT will just transform
the imaginary part
of a subluminal geometric object into the real part of the transformed superluminal object, and vice versa,
in the simple typical case of the
Lorentz transformation ``dual" of the identity transformation ($v=0$) and therefore
corresponding to $V=\infty$ as provided by the standard duality relation $V \equiv 1/v$.  Remember[14] that in NRR only
the speed $c$ of light is invariant (and not the infinite speed!)

\h In conclusion, according to NNR, an imaginary unit, besides a double sign, must formally enter in the transverse components
of Eqs.24 and 25 in Ref.[1]. A mathematically sound way for expressing the SLTs is the only
(formal) problem left with NNR, and we are glad of the renewed interest in it with the hope that
some scholar will tackle it; one of the authors (ER) and his collaborators have not succeeded yet in finding out a really
satisfactory formal expression, even by using Clifford algebras (unless one goes from four to six dimensions). [A similar
probem is actually met also when passing on
from the exterior to the interior of a black-hole (BH), using not $\rho$ and $t$ only, but four coordinates;
and even more in the
case of a non-spherically symmetric BH].

\

\h {\em Nevertheless,} one can usefully apply our SLTs. As realized also by Seshadri[1],
in the particular case of the complete Maxwell equations, which are covariant under
rotations in the 6D space ($\Ebf, \Hbf$) and in particular under the transformations $\Ebf \rightarrow
i \; \Hbf; \ \ \ \Hbf \rightarrow -i \; \Ebf$, one can obtain the generalized Maxwell equations[14,28]
by meaningfully using the SLTs in their present form (given for instance in
pages 123--126 of Ref.[14]). Our generalized Maxwell equations are presented in Sec.6.15,
and particularly in Sec.15 of Ref.[14] and references therein (for example, see Ref.[28]).

\

\h {\em Even more,} one can have useful recourse to our SLTs, merely by remembering that they {\em change sign to the
quadratic forms} \
$x_\mu x^\mu$,  \ $p_\mu p^\mu$,  and
$x_\mu p^\mu$. \ This has been done in Refs.[18,14] to get, e.g., the result shown in our Fig.2. \ Indeed, to get the
shape of a simple tachyon (the transform of a quite simple, ball-shaped bradyon moving in the $x$-direction),
one starts with the world-tube
associated in spacetime with the considered, spherical, subluminal object. For subluminal speeds $v$, the world-tube
will be inclined with respect to the $t$-axis of an angle $|\alpha|< 45^{\rm o}$; and, by cutting it with a plane
$t$ = constant, one gets of course the ellipsoid produced by the Lorentz space-contraction along $x$.  In the case of
a superluminal speed $V$, the world-tube happens to be outside the light cone,where the geometry is pseudo-Euclidean.
Therefore, we wrote down in manifestly covarint form the equation of the cylindrical surface of the subluminal tube, and
transformed it superluminally: That is to say, by changing sign to the quadratic forms entering the said equation.
Then, by cutting with  $t$ = constant the resulting new equation, we easily obtained[18,14] the shape of the
corresponding superluminal object, which resulted to be the X-shaped region included between an indefinite double-cone
and a two-sheeted hyperboloid (see Fig.2).

\

\

{\bf 2.5 \ Reply to Seshadri's Section 6}

\

In Sec.6 of Ref.[1], our publication [7] is re-examined. In that paper, we just investigated the
toy-model of a point-like superluminal charge (a very bad approximation in all cases, and particularly in
the tachyonic case, always leading as we know to divergencies), and we carefully mentioned that in such a case one
{\em has} to meet singularities on the double cone. The corresponding claims by Seshadri just
confirm our own statements[7]. Of course, for the reasons previously discussed, calculations suitably
performed in order to represent half retarded and half
advanced fields yield a double-cone geometry.

\

\

\

{\bf 3. \ Acknowledgements}

\

The authors are grateful to two anonymous Referees for some very useful comments.

\

\

{\Large{ \bf 3. \ Bibliography}}

\

[1] S.R.Seshadri, ``Cherenkov radiation versus X-shaped localized waves:
Comment," J. Opt. Soc. Am. A {\bf 29}(12), 2532-2535 (2012): previous article.

[2] M.Zamboni-Rached, E.Recami and I.M.Besieris, ``Cherenkov radiation versus
X-shaped localized waves,"
J. Opt. Soc. Am. A {\bf 27}, 928-934 (2010).

[3] S.C.Walker and W.A.Kuperman, ``Cherenkov-Vavilov
formulation of X waves,"
Phys. Rev. Lett. {\bf 99}, 244802 [four pages] (2007).

[4] {\em Localized Waves, Theory and Applications}, ed. by
H.E.Hern\'andez-Figueroa, M.Zamboni-Rached, and E.Recami (J.Wiley;
New York, 2008).

[5] J.-y. Lu and J.F.Greenleaf, ``Experimental verification of
nondiffracting X-waves,"
IEEE Trans. Ultrason. Ferroelec. Freq. Control {\bf 39}, 441-446 (1992).

[6] P.Saari and K.Reivelt, ``Evidence of X-shaped propagation invariant
localized light waves,"
Phys. Rev. Lett. {\bf 79}, 4135-4138 (1997).

[7] E.Recami, M.Zamboni-Rached and C.A.Dartora,
``Localized X-shaped field generated by a superluminal charge,"
Phys. Rev. E {\bf 69}, 027602 [four pages] (2004); and refs. therein.

[8] E.Recami and M.Zamboni-Rached: ``Localized Waves: A Review",
Advances in Imaging \& Electron Physics (AIEP) {\bf 156}, 235-355
(2009) [121 printed pages).

[9] A.Sommerfeld, ``\"Uberlichtgeschwindigkeitsteilchen,"
K. Ned. Akad. Wet. {\bf 8}, 346 (1904).

[10] A.Sommerfeld, ``Zur Electronentheorie (3 Tiele)," Nach. Kgl. Ges. Wiss. G\"ottingen,
Math. Naturwiss. Klasse 99-130, 363-439 (1904), 201-236 (1905).

[11] P.O. Fr\"oman, ``Historical background of the tachyon concept,"
Arch. Hist. Exact Sci. {\bf 48}, 373-380 (1994); DOI: 10.1007/BF00375087.

[12] O.-M.Bilaniuk, V.K.Deshpande and E.C.G.Sudarshan, `` `Meta' relativity,"
Am. J. Phys. {\bf 30}, 718-723 (1962).

[13] O.-M.Bilaniuk and E.C.G.Sudarshan, ``Particles beyond the light barrier,"
Phys. Today {\bf 22}, 331-339 (1969).

[14] E.Recami, ``Classical theory of tachyons,"
Rivista Nuovo Cim. {\bf 9}(6), 1-178 (1986), monographic issue no.6.

[15] R.Mignani and E.Recami: ``Crossing Relations Derived from
(Extended) Relativity",
Int. J. Theor. Phys. {\bf 12}, 299-320 (1975).

[16] M.Pav\v{s}i\v{c} and E.Recami: ``Charge Conjugation and Internal
Space-Time Symmetries",
Lett. Nuovo Cim. {\bf 34}, 357-362 (1982).

[17] E.Recami: ``Tachyon Mechanics and Causality; A Systematic
Thorough Analysis of the Tachyon Causal Paradoxes,"
Found. Phys. {\bf 17}, 239-296 (1987).

[18] A.O.Barut, G.D.Maccarrone and E.Recami: ``On the shape of tachyons,"
Nuovo Cimento A {\bf 71}, 509-533 (1982).

[19] R.Mignani and E.Recami: ``Tachyons do not emit Cherenkov
radiation in vacuum,"
Lett. Nuovo Cim. {\bf 7}, 388-390 (1973).

[20] A.B.Utkin, ``Droplet-shaped waves: causal finite-support analogs of
X-shaped waves,"
J. Opt. Soc. Am. A {\bf 29}, 457-462 (2012).

[21] P.M.Morse, {\em Theoretical Acoustics}
(Princeton Univ. Press, 1985).

[22] E.Arias, C.H.G.Bessa and N.F.Svaiter, ``An analog
fluid model for some tachyonic effects in Field Theory,"
Mod. Phys. Lett. A {\bf 26}, 2335-2344 (2011); and refs. tehrein.

[23] E.Recami: ``The Tolman antitelephone paradox: Its solution
by tachyon mechanics," 1985, reprinted in
Electronic J. Theor. Phys. (EJTP) {\bf 6}(21), 1-8 (2009).

[24] E.Recami: ``Superluminal motions? A bird's-eye view of the experimental
status-of-the-art",
Found. Phys. {\bf 31}, 1119-1135 (2001).

[25] E.Recami: ``Superluminal waves and objects: An up-dated
overview of the relevant experiments",
arXiv:0804.1502 [physics].

[26] E.Recami and  W.A.Rodrigues: ``A model theory for tachyons in two
dimensions," in {\em Gravitational Radiation and Relativity} (vol.3
of the Proceedings of the Sir Arthur Eddington Centenary Symposium),
ed. by J.Weber and T.M.Karade (World Scientific; Singapore, 1985),
pp.151-203.

[27] A.O.Barut and Chandola: ``Localized' tachyonic wavelet solutions
to the wave equation,"
Phys. Lett. A {\bf 180}, 5-8 (1993).

[28] E.Recami and R.Mignani: ``Magnetic Monopoles and Tachyons in
Special Relativity,"
Phys. Lett. B {\bf 62}, 41-43 (1976).


\end{document}